\documentclass[a4paper]{jpconf}
\usepackage{graphicx}
\usepackage{nicefrac}
\usepackage{enumitem}
\usepackage{amsmath,amssymb,mathrsfs}

\begin{document}
\title{The attraction between\\ antiferromagnetic quantum vortices \\
 as origin of superconductivity in cuprates}

\author{P. A. Marchetti$^*$}

\address{Dipartimento di Fisica e Astronomia, Universit\'a di Padova, INFN,\\
Padova, I-35131,Italy\\
$^*$E-mail: marchetti@pd.infn.it}

\begin{abstract}
We propose as key of superconductivity in (hole-doped) cuprates a novel excitation of magnetic origin, characteristic of two-dimensions and of purely quantum nature: the antiferromagnetic spin vortices. In this formalism the charge pairing arises from a Kosterlitz-Thouless-like attraction between such vortices centered on opposite N\'{e}el sublattices. This charge pairing induces also the spin pairing through the action of a gauge force generated by the no-double occupation constraint imposed in the $t$-$J$ model of the CuO planes of the cuprates. Superconductivity arises from coherence of pairs of excitations describing Zhang-Rice singlets and it is not of standard BCS type. We show that many experimental features of the cuprates can find a natural explanation in this formalism.
\end{abstract}



\section{Introduction}\label{aba:sec1}
Thirty years after the discovery of the first high-$T_c$ superconducting cuprate \cite{bedmu}, the microscopical mechanism behind superconductivity in this class of materials is still not understood, despite constant experimental advances. It is commonly believed that antiferromagnetism (AF) is a key ingredient for the superconductivity in cuprates, then a natural pairing glue would be provided by the spin fluctuations, i.e. antiferromagnetic spin-waves (see e.g. Ref. \cite{manske}). Their action would be enhanced by nesting of the Fermi surface (FS), but evidence for this is not so clear.
We propose as pairing glue another
excitation still emerging from AF, but of
purely quantum origin: antiferromagnetic spin vortices.
In the antiferromagnetic phases  the spin group $SU(2)$ is broken to $U(1)$, the quotient $SU(2)/U(1)$ is isomorphic to the 2-sphere $S^2$ whose points label the directions of the magnetization. Their fluctuations are described by spin waves. The unbroken $U(1)$ group describes unphysical gauge fluctuations. However in two dimensions (2D) one can consider vortices of Aharonov-Bohm type in this $U(1)$; due to AF the vortices have opposite chirality when centered in two different N\'{e}el sublattices, hence we dub them antiferromagnetic spin vortices.
Lowering the temperature such gas of vortices in 2D undergoes a Kosterlitz-Thouless-like transition, with the formation of a finite density of vortex-antivortex pairs.
If the vortices are centered on charges, this induces a new form of charge-pairing, again due to AF, but different from the spin-fluctuation pairing. As discussed later this pairing finally leads to superconductivity.

 To present how this is realized in the (hole-doped) cuprates is the fundamental aim of this paper. One discovers that through this key idea many structural features of the phase diagram of the cuprates could be understood and many physical properties successfully computed. The derivations are based on well-defined conjectures and approximations and many experimental consequences are consistent with availabla data, as sketched in the final section.  Let us stress that, due to their structure, the antiferromagnetic vortices are specific to quasi-2D doped antiferromagnets and some phenomena they give rise to, like the charge pairing discussed above, the induced spin-pairing and the metal-insulator crossover of in-plane resistivity, are peculiar to this approach and do not appear in this form in other approaches to the physics of cuprates.
 
Whenever we succeed we present the intuitive ideas at the beginning of the section, giving later on a brief summary of their formal implementation and deferring to the references for the explicit proofs.

This paper reviews the results on a mechanism for superconductivity arising from a long joint project  on a spin-charge gauge approach to the physics of cuprates initiated with Z.-B. Su and L. Yu, to whom I express a deep gratitude for all the knowledge and ideas shared with me. Along the way we profit of fundamental contributions from many researcher, among which a leading role have been played by F. Ye and G. Bighin, but it is also a pleasure to cite as co-authors J.-H. Dai,  L. De Leo, G. Orso, A. Ambrosetti, M. Gambaccini and last but not least to gratefully acknowledge the original insight by J. Froehlich.

\section{Model and phase diagram}\label{aba:sec2}

 The high $T_c$ cuprates share a layered structure incorporating one or more copper–oxygen (CuO) planes; an excellent recent summary of their properties can be found in Ref. \cite{uchida}. There is no consensus
yet on the theoretical interpretation of their low-energy physics, but 
an agreement has been achieved that superconductivity is due to formation of Cooper pairs with principal locus the CuO planes and that 
the order parameter is a spin singlet with orbital symmetry $d_{x^2-y^2}$.
Still a large consensus have the ideas that
the formation of Cooper pairs takes place independently within different
CuO multilayers and that repulsive electron-electron interaction plays a key role, the electron-phonon interaction being not the principal mechanism of their formation.
However, even with these restrictions several quite different proposal for the pairing mechanism leading to superconductivity have been made. A recent, partial comparison of different theories with some experimental data can be found in Ref. \cite{bok}.

Most of the researchers believe that for the hole-doped cuprates, considered in this paper, the key actors in the superconducting transition are the so-called Zhang-Rice singlets \cite{zr}. Let us sketch what they are and why they naturally lead to a $t$-$J$ model-like description of the CuO planes. This basic structural unit is formed
by a square lattice of Cu atoms with oxygen atoms in the middle of the sides of the square. In the undoped materials the relevant orbitals are the $3d_{x^2-y^2}$ orbitals of the Cu, containing one electron per site with a large Coulomb repulsion prohibiting a low-energy double occupation, and the $p$ orbitals of the O, oriented along the sides of the square and completely filled. The spins of the Cu electrons at low temperature are antiferromagnetically ordered.  The Cu and O orbitals have similar energies and they strongly hybridize. If a doping hole is introduced, it goes predominantly in the combinations of four oxygen $p$ orbitals centered around a copper site forming a spin-singlet with the spin of the corresponding Cu site. These are the Zhang-Rice singlets and, from the point of view of the square lattice of the Cu, the corresponding site has 0 charge and 0 spin. One can thus describe effectively the low-energy of the CuO plane in terms of a square lattice, corresponding to the positions of Cu atoms, whose sites are either singly occupied or empty. Furthermore every oxygen orbital contributing to the hybridization is shared by two Cu, so that these singlets, described by empty sites, can hop between the sites of the above square lattice. Therefore a reasonable Hamiltonian $H$ describing the low-energy physics of the CuO planes is given by the $t$-$J$ model
\begin{equation}
\label{eq1}
H_{t-J}=\sum_{\langle i,j \rangle} P_G \left[ -t c^*_{i \alpha} c_{j \alpha} + J {\vec S}_i \cdot {\vec S}_j \right] P_G,
\end{equation}
 where $i,j$ denote nearest neighbour (nn) sites of the lattice, $c$ the hole field operator, $P_G$  the Gutzwiller projection eliminating double occupation, $\alpha$ the spin indices and  summation over repeated spin (and vector) indices is understood, here and in the following. A typical value for the nn hopping is $t \approx 0.3 \hspace{2pt} \mathrm{eV}$ and for the anti-ferromagnetic Heisenberg coupling $J \approx 0.1 \hspace{2pt} \mathrm{eV}$. (These are the values used in the numerical computations reported in the last section.) Furthermore to get a reasonable shape of the Fermi surface in the tight binding approximation one adds to the nn hopping term at least a nnn term with coefficient $t'$, $|t'| \sim 0.03 - 0.1 \hspace{2pt} \mathrm{eV}$, strongly material dependent. 
 
Let us briefly sketch the structure of the phase diagram of the (hole-doped) cuprates;
 for a very recent review see Ref. \cite{hussey}.
 In the ground state of the undoped materials all the Cu sites are singly occupied, thus realizing an antiferromagnetic Mott insulator, due to the strong on-site Coulomb repulsion. Long-range antiferromagnetism disappears when the doping concentration, $\delta$, in a CuO plane exceeds a critical value ( $\delta \approx 0.03$ at $T=0$), through a N\'{e}el transition, but still a ``charge-transfer'' gap exists between the Zhang-Rice band and the upper Hubbard band of the Cu. Superconductivity (SC) appears when $\delta$ reaches $\approx 0.05$ and as a function of doping the superconducting temperature $T_c$ exhibits a dome shape. The doping concentration corresponding to the highest $T_c$ is called optimal doping; the cuprates with lower doping are called underdoped and those with a higher one overdoped. In a V-shaped region in  the $\delta$-$T$ plane around $\delta \approx 0.19$,  called ``Strange Metal'' (SM), ARPES experiments reveal a Fermi surface roughly consistent with the band calculations, but the quasi-particle peak is broad and the incoherent tail anomalously large. Also the transport presents strange features, like the linear in $T$ behaviour of in-plane resistivity. In the underdoping region above $T_c$ and
below a temperature $T^*$ indicating a crossover (or transition) to the SM there is a region called Pseudogap (PG), characterized by a suppressed density of states near the Fermi level. The current picture of the pseudogap opening is the 
following (see e.g. Ref.\cite{yoshida}): in the large Fermi surface seen in the SM
as the temperature decreases below $T^*$, the pseudogap first opens near
(0,$\pm \pi$) and $(\pm \pi, 0)$ (antinodal region), it then gradually ``eats'' up the original FS,
converting it into a ``pseudogapped'' part, eventually leaving only
short disconnected arcs around ($\pm \pi/2,\pm \pi/2$) (nodal region). Finally, these
arcs shrink abruptly to nodal points of the superconducting gap function at the superconducting transition, maintaining the $d$-wave
symmetry. 
Actually some experimentalists and theorists suggest the existence of two ``pseudogap temperatures'' (see e.g. Refs. \cite{ho},\cite{ba}), a ``high'' one that joins the superconducting dome in the overdoped side and a ``low'' one crossing the superconducting dome and reaching $T=0$ near $\delta \approx 0.19$ if superconductivity is suppressed. The first one corresponds to the emergence of the ``pseudogap'' phenomenology on the FS described above, but also e.g. to a decrease of the Knight shift or a deviation from linearity of in-plane resistivity. The second one corresponds e.g. to an inflection point of the resistivity curve and to a broad peak in the specific heat coefficient.
Another crossover line dome-shaped above the dome of the superconducting transition signals the onset of a Nernst signal due to magnetic vortices and  diamagnetism \cite{liwan}. We call ``Nernst'' (N) the non-superconducting region below this crossover. Both the ``low'' pseudogap and the Nernst crossover line seem to be universal in the cuprates, whereas somewhat material-dependent appear the ``high'' pseudogap crossover and even more the superconducting transition.
Finally, there is a crossover between AF and SC below which there is a "spin-glass" region and recently it has been firmly established \cite{bad} a dome-shaped crossover, quite distinct from the superconducting dome, around $\delta \approx 1/8$ below which one finds evidence of charge-density waves.

\section{Hints from 1D}\label{aba:sec3}

Two natural questions arising from the previous sections are: how the antiferromagnetic vortices appear in the $t$-$J$ model and how much they explain of the above sketched phenomenology of the cuprates.
Let us come back to the $t$-$J$ model. Inspired by an idea pioneered by Anderson \cite{an} and Kivelson \cite{kiv}, one can get rid of the no-double occupation constraint imposed by the Gutzwiller projector $P_G$ by rewriting the hole field $ c_{\alpha}$ as a product of a charge 1 spinless fermion field $h$, the holon, and a neutral spin 1/2 boson field $\tilde s_\alpha$, the spinon, as $c_\alpha=h^* \tilde s_\alpha$. (We introduced the tilde because we will denote by $s_\alpha$ a slightly different field). The holon $h$ being spinless implements exactly the Gutzwiller constraint by the Pauli principle. Furthermore if the constraint $ \tilde s^*_\alpha \tilde s_\alpha=1$ is imposed, since  $ c^*_\alpha c_\alpha=1-h^* h$ we see that $h^* h$ is just the density of empty sites in the model. However, if we treat the holon in mean-field, precisely because it is spinless one doesn't get a reasonable Fermi surface. Therefore we need to ``dress'' it, and actually we will see that also the spinon needs a ``dress''. A hint on how to proceed comes from the one-dimensional $t$-$J$ model, which in some limit is exactly solvable, thus allowing a check.
Let us here give the intuitive idea, then we briefly sketch its formal implementation in the next section.

Consider the Heisenberg spin 1/2 chain, describing the 1D $t$-$J$ model in the limit of zero doping and take as reference state that with the spin antiferromagnetically ordered, the semi-classical ground state mimicking the N\'{e}el order in 2D. Now let us insert a dopant by removing the spin from a site; as a consequence the two neighbouring spins will be ferromagnetically aligned. Let then the empty site to hop by a simultaneous opposite hopping of the spin, then we get two separate excitations. There is an empty site, but with neighbouring spins antiferromagnetically aligned, thus carrying charge but not spin; the corresponding excitation is the holon. There is another site where there is a domain wall between two different N\'{e}el sublattices, hence carrying spin 1/2 but neutral; the corresponding excitation is the spinon (see e.g. Ref. \cite{giama}). Notice however that attached to the site with a spin mismatch of 1/2 there is a string of spins flipped w.r.t. the reference state from that site to the holon position, which is integral part of the spinon excitation. It turns out that, due to this ``spin string'', interchanging two spinon fields one gets a phase factor $e^{\pm i \pi/2}$, hence the spinon is a semion. A semion, in fact, is  
a particle excitations obeying braid statistics, which can be characterized precisely by the phase factor $e^{\pm i \pi/2}$ acquired by the many-body wave-function or by the product of two fields operators when two semions are exchanged (see e.g. Ref. \cite{wil}), the sign depending on the relative orientation of the exchange w.r.t. the space.
 Naively one can describe the holon as a spinless fermion $h$, but to maintain the fermion statistics of the hole one must add a ``charge string'' turning it into a semion too.
The charge string has an additional effect: to modify the Haldane (exclusion) statistics \cite{ha} of the holons assigning to them an exclusion statistics parameter 1/2. A way of characterizing the exclusion statistics parameter $g$ is that, setting the Fermi energy, the maximal particle number $N_g$ for given $g$ is determined by $N_0=N_g(1-g)$, where $N_0$ is the maximal fermion filling number for $g=0$, the exclusion statistics parameter for fermions \cite{wu}. Therefore for the holon
the maximum occupation at fixed momentum is 2 and the Fermi momenta of the spinless semionic holon is the same of the Fermi momenta of a spin 1/2 fermion.
Hence, since the spinon has no chemical potential, the composite hole, product of spinon and holon, satisfies the Luttinger theorem. 
This was exactly what we were looking for in 2D. Therefore, if we understand how to implement mathematically this picture, this will be a guide for tackling the 2D problem. Although the unphysical mathematical detour in 1D of the next section will be somewhat long, we think it is useful since many ideas there can be explicitly tested by  comparison with known results and they will be taken up for the 2D treatment.

\section{The spin-charge gauge formalism}\label{aba:sec4}

Let us sketch how the  picture described in the previous section can be implemented mathematically in the path-integral formalism using Chern-Simons gauge fields, in the so-called spin-charge gauge approach; for details see Ref. \cite{msy}. We start coming back to 2D and basing
our theoretical treatment of the $t$-$J$ model on the
following
\cite{fro}\cite{np}

We embed the lattice of the 2D $t$-$J$ model in a 2-dimensional space, denoting by $x=(x^0,x^1,x^2)$ the coordinates of the corresponding 2+1 space-time, $x^0$ being the euclidean time. We couple the fermions of the $t$-$J$ model to a $U(1)$ gauge field,
$B^\mu$, gauging the global charge symmetry, and to an $SU(2)$ gauge field,
$V^\mu$, gauging the global spin symmetry of the model, and we assume that
the dynamics of the gauge fields is described by the Chern-Simons actions:
\begin{displaymath}
   S_{c.s.} (B) = - \frac{1} {2 \pi} \int d^3 x \epsilon_{\mu\nu\rho}
   B^\mu \partial^\nu B^\rho (x),
\end{displaymath}
\begin{equation}
   S_{c.s.} (V) = \frac{1} {4\pi} \int d^3 x {\rm Tr} \epsilon_{\mu\nu\rho}
   [V^\mu \partial^\nu V^\rho + {2\over 3} V^\mu V^\nu V^\rho](x),
\label{eq2}   
\end{equation}
where $\epsilon_{\mu\nu\rho}$ is the Levi-Civita anti-symmetric tensor in 3D.
Then the spin-charge (or $SU(2) \times U(1)$) gauged model so
obtained is exactly equivalent to the original $t$-$J$ model.
In particular the spin and charge invariant correlation functions of the fermions fields  $c_{j \alpha}$ of the $t$-$J$ model are exactly equal to the correlation functions of the fields $ \exp[i \int_{\gamma_j} B] P(\exp[i \int_{\gamma_j} V])_{\alpha \beta} c_{j \beta}$ , where $c$ denotes now the fermion field of the gauged model, ${\gamma_j}$ a string at constant euclidean time connecting the point $j$ to infinity and $P(\cdot)$ the path-ordering, which amounts to the usual time ordering $T(\cdot)$, when ``time'' is used to parametrize the curve along which one integrates.
(For a careful discussion of boundary conditions and further details, see Ref. \cite{fro}.)

A good feature
of introducing the above gauge fields is that they allow
a more flexible treatment of charge and spin responses
within a spin-charge decomposition scheme; we have argued above that spin and charge are quite independent in 1D and this turns out to be partially true even in 2D. A comment on the notations: for lattice fields space coordinates  are denoted by sub-indices and time coordinate as a variable (e.g. $c_j(x^0)$), but often omitted for simplicity; for continuum fields space-time coordinates are denoted as variables without indices and often space coordinates with the vector symbol (e.g. $x=(x^0, \vec x)$).

Let us give the key ideas of the proof of the above theorem for the
partition function. We represent the partition function of the gauged
model in the first-quantized formalism in terms of the worldlines
of  fermions. After integrating out the gauge fields, due to the chosen coefficients of the Chern-Simon action, the effect
of the coupling to $B_\mu(V_\mu)$ is only to give a factor $e^{\pm i
{\pi\over 2}}(e^{\mp i{\pi\over 2}})$  for any single exchange of the
fermion worldlines, so  the two effects cancel each other exactly. The two signs in the phases correspond to the two possibilities of under-crossing or over-crossing of the worldlines; due to the Gutzwiller projection there are no intersections between the worldlines, so the crossing are well-defined.
 We just mention that the role of the strings $\gamma$, appearing in the representation of the fields in the theorem, is to ``close'' the worldlines emerging when a fermion is created and annihilated, to guarantee gauge-invariance also in the correlators.
 
To discuss the $t$-$J$ model in 1D we simply restrict the lattice to 1D and choose the strings $\gamma$ parallel to the lattice, whose space direction we label with the index 1, towards $+\infty$ but infinitesimally shifted in the orthogonal direction, labelled by the index 2, to avoid the intersections with the worldlines of the fermions.

 We apply to the fermion of the gauged model the spin-charge decomposition discussed in Sec. 3, but now we identify $ \exp[i \int_{\gamma_j} B]h_j$ as the holon and $ P(\exp[i \int_{\gamma_j} V])_{\alpha \beta} \tilde s_{j \beta}$ as the spinon fields. The Chern-Simons coupling automatically ensures that the corresponding field operators obey both semionic braid statistics. As we will see the two strings introduced above are precisely the ``dressing'' alluded in Sec. 2, reproducing in fact the correct result in 1D.

To understand this result  we start noticing that in absence of holons the spin of the Heisenberg model is written in terms of $\tilde s_\alpha$ as $\vec S= \frac{1}{ 2}\tilde s^*_\alpha \vec \sigma_{\alpha\beta} \tilde s_\beta$, where $\vec \sigma$ denote the Pauli matrices. Therefore, to come closer to the picture discussed in the previous section we first gauge-fix the $SU(2)$ gauge invariance imposing (even in the presence of holons) the requirement that the spins are antiferromagnetically ordered, by setting
\begin{eqnarray}
\label{eq3}
\tilde{s}_j = \sigma_x^{|j|} \begin{pmatrix}
 1\\0 
\end{pmatrix},
\end{eqnarray} 
where $|x|=x^1+x^2$. (In 1D for $j$ a lattice site we have simply $|j|=j$, but this notation is useful in 2D.)
 
After the holon-spinon decomposition introduced above the action of the gauged $t$-$J$ model at doping $\delta$ is given by \cite{msy98} :
\begin{eqnarray}
\label{eq4} 
&S(h,h^*,B,V)=\int_0^\beta dx^0 \sum_j h^*_j [\partial_0 -i B_0(j)-\delta] h_j + i B_0(j)+i(1-\\& h^*_j h_j)( \sigma_x^{|j|} V_0(j)  \sigma_x^{|j|})_{11} - \sum_{\langle i,j \rangle} t h^*_j 
 \exp[i \int_{\langle i,j \rangle} B]h_i (\sigma_x^{|i|} P(\exp[i \int_{\langle i,j \rangle}  V])\sigma_x^{|j|})_{11}]\nonumber\\& +\frac{J}{2}(1-h^*_jh_j)(1-h^*_ih_i) [|(\sigma_x^{|i|} P(\exp[i \int_{\langle i,j \rangle}  V])\sigma_x^{|j|})_{11}|^2-\frac{1}{2} ]+S_{c.s.}(B)+S_{c.s.}(V).\nonumber
\end{eqnarray} 
The two-point correlation function of the hole can be written as:
\begin{eqnarray}
\label{eq5}
&\langle c^*_{x\alpha}(x^0) c_{y\alpha}(y^0) \rangle =\int \mathcal{D}h\mathcal{D}h^*\mathcal{D}B \mathcal{D}V  \exp[-S(h,h^*,B,V)] h^*_x \exp[-i \int_{\gamma_x} B](x^0)\nonumber\\ &\exp[i \int_{\gamma_y} B]h_y(y^0)(\sigma_x^{|x|}P(\exp[i \int_{\gamma_x} V]))_{1\alpha}(x^0)(P(\exp[i \int_{\gamma_y} V])\sigma_x^{|y|})_{\alpha1}(y^0) \nonumber\\
&[ \int \mathcal{D}h\mathcal{D}h^*\mathcal{D}B \mathcal{D}V  \exp[-S(h,h^*,B,V)]]^{-1},
\end{eqnarray} 
where $x,y$ are lattice sites and  $x^0,y^0$ are euclidean times. 

Let us discuss the case of 1D. Since the 0-component of $B$ appears linearly in (\ref{eq5}) , we can safely integrate it out getting for the numerator of (\ref{eq5}) the constraint (with $\mu,\nu=1,2, z \in \mathbf{R}^3$):
\begin{eqnarray}
\label{eq6}
\epsilon_{\mu\nu}\partial^\mu B^\nu(z)=\pi \delta (z^2) [\sum_j \delta(z^1-j)(1-h^*_jh_j)(z^0)].
\end{eqnarray}
By imposing the gauge-fixing $B^2=0$ one finally gets
\begin{eqnarray}
\label{eq7}
B^1(z)=\frac{\pi}{2} {\rm sgn}(z^2) [\sum_j \delta(z^1-j) (1-h^*_jh_j)(z^0))],
\end{eqnarray}
so that the holon field turns out to be given by
\begin{eqnarray}
\label{eq8}
  \exp[i \int_{\gamma_j} B]h_j=\exp[i \frac{\pi}{2}\sum_{\ell>j} (1-h^*_\ell h_\ell)] h_j. 
\end{eqnarray}
If the doping is $\delta$, the fermion $c_\alpha$ of the 1D $t$-$J$ model in the tight-binding approximation has a Fermi momentum $ \pi(1-\delta)/2$ since two fermions with opposite spin can have the same momenta. For the spinless fermion $h^*$ the Fermi momentum would be $ \pi(1-\delta) $ since only one spinless fermion con have a fixed momenta. However, if we consider the phase  string attached to the spinless fermion in (\ref{eq8}) we see that, since the expectation value of $h^*_\ell h_\ell$ is $\delta$, it contributes to the Fermi momentum a term $-\pi(1-\delta)/2$, so that the semionic holon of (\ref{eq8}) obeys an Haldane statistics of parameter 1/2. Hence when the electron is reconstructed combining the holon with the spinon, it has the same Fermi surface (in the leading approximation) of the tight-binding approximation for the fermion.  With the due changes the same mechanism will be advocated in 2D.

The treatment of the spinon is slightly more involved. Since we have already gauge-fixed the $SU(2)$ gauge-invariance in (\ref{eq3}), the gauge field $V$ has to be integrated without gauge-fixing. Then we split the integration over $V$ into
an integration over a field $\bar V$, satisfying the gauge-fixing condition $\bar V^2=0$
and its gauge transformations expressed in terms of an
$SU(2)$-valued scalar field $g$,
i.e., $V^\mu= g^\dagger \bar V^\mu g+g^\dagger\partial^\mu g$. Notice that $ P(\exp[i \int_x^y V])=g^\dagger_y P(\exp[i \int_x^y \bar V])g_x$.
We now find the configuration of $g$ that optimize the partition function of holons in a fixed, but holon-dependent, background.
It is rigorously proved in Ref. \cite{np} that such configuration is given by $g^m_j=\exp[ -i \frac{\pi}{2} \sigma_x \sum_{\ell>j} h^*_\ell h_\ell ]$. This is precisely the ``string'' of spin flipping that we have encountered in the preliminary qualitative discussion in the previous section. Finally we set $g=Ug^m$, with $U \in SU(2)$ describing the fluctuations around the optimal configuration. We write 

\begin{eqnarray} 
\label{eq9}
U= \begin{pmatrix}
 s_1 &-s_2^*\\ s_2 & s_1^*
\end{pmatrix},
\end{eqnarray} 
and we will call in the following the $s_\alpha$ again ``spinons''. One such spinon can be identified with the 1/2-spin local configuration in the previous section.
As a ``mean field'' approximation we neglect the fluctuations $U$ in the calculation of $\bar V$. Proceeding now as done for the $B$ field and integrating over  $\bar V^0$ one finds that $\bar V^1$ is diagonal and one of the two components of $P(\exp[i \int \bar V])$ acts erasing the pure phase factor of  $g^m$ , the other acts doubling this phase factor. One can then show that in this way one can reproduce \cite{np} the correct longwavelength limit of the correlation functions of the 1D $t$-$J$ .

Let us remark an interesting feature \cite{msy} of the above approach: thanks to the holon-depending spin flips, since the holon in a oriented hopping link is at its end,  $(\sigma_x^{|i|} P(\exp[i \int_{\langle i,j \rangle}  V]\sigma_x^{|j|})_{11}]$ in the $t$-term of (\ref{eq4}) equals $s^*_{\alpha i}s_{\alpha j}$, whereas in the $J$-term, where there are no holons, it equals  $\epsilon_{\alpha \beta} s_{\alpha i}s_{\beta j}$. But under the constraint  $s^*_{\alpha j}s_{\alpha j}=1$ the following identity holds: $|s^*_{\alpha i}s_{\alpha j}|^2+|\epsilon_{\alpha \beta} s_{\alpha i}s_{\beta j}|^2=1$ so that if one optimizes the $t$-term choosing $|s^*_{\alpha i}s_{\alpha j}|=1$, simultaneously one optimizes also the $J$-term. This optimizing property is somewhat strange, since the same expression in terms of $V$ gives rise to different expressions in terms of the spinons $s$ in the $t$- and the $J$-terms. It is intrinsically due to the $SU(2)$ gauge degrees of freedom, absent in the more standard gauge approach of the slave boson formalism, which involves only $U(1)$ gauge degrees of freedom.
Let us finally briefly comment on why in the picture of Sec. 3 the position of the spinon, corresponding to a spin kink, is different from that of the holon. The reason is that if we eliminate the holons in the remaining Heisenberg chain there is a gas of spin kink-antikinks that interpolate between two semiclassical realizations of the N\'{e}el order, related by parity in 1D. However, since as well known there is no N\'{e}el order in 1D, this gas is dense and e.g. an  anti-kink of this gas annihilate a spin kink generated by the holon leaving a spin kink arbitrarily far from the holon position.
Carrying the know-how gained in 1D we come back now to the physical 2D model.

\section{The slave-particle gauge field}\label{aba:sec5}

Let us try naively to export to 2D the string mechanism discussed in 1D in Sec. 3. We immediately find a big difference: when there is a string of spin flips in a chain the two adjacent chains will have the spin with the same orientation of those of the string and being the spin interaction antiferromagnetic this configuration costs an energy proportional to the length of the string, i.e. holon and spinon connected by a string of spin flips are confined! Here we see the appearance of a new actor that in our discussion in 1D we have overlooked: a gauge force between spinon and holon.
Mathematically its origin comes from the spin-charge decomposition of the fermion of the (gauged) $t$-$J$ model: $c_{j \alpha}= \exp[i \int_{\gamma_j} B]h_j P(\exp[-i \int_{\gamma_j} V])_{\alpha \beta} \tilde s^*_{j \beta}$. This decomposition is clearly invariant under the $U(1)$ gauge transformation: $h_j \longrightarrow h_j e^{i \Lambda_j}, \tilde s_{j \alpha} \longrightarrow  s_{j \alpha} e^{i \Lambda_j}$ (we can also replace $\tilde s$ by $s$) , with $\Lambda$ a real lattice function. We call this emergent symmetry slave-particle (or h/s) gauge symmetry, not to be confused with the charge gauge symmetry whose gauge field is $B$. At least in the long wavelength continuum limit one can make this symmetry explicit by introducing the related gauge field that we denote by $a^\mu$. This can be done in absence of holons simply by assuming a continuum limit of the spinon field $s$ of the form
$s_{x \alpha}(x^0) \longrightarrow s_\alpha(x) + (-1)^{|x|} p_\alpha(x)$, where $s$ and $p$ in the r.h.s. are continuum fields. Integrating out the ferromagnetic component $p$, as shown in Ref.\cite{sac}, both in 1D and 2D we obtain a low-energy spinon Lagrangian in the form of a $CP^1$ model: 
\begin{eqnarray}
    \mathcal{L}_s=   \frac{1}{g}
   \left[ v_{s}^{-2} \left| \left( \partial_{0} - i a_{0} \right) s_{\alpha}
   \right|^{2} + \left| \left( \vec \nabla - i \vec a
   \right) s_{\alpha} \right|^{2} 
   \right],
   \label{eq10}
\end{eqnarray}
where  $g \sim \epsilon^{D-1} J^{-1}$ and $v_s \sim J \epsilon$ is the spinon velocity, with $\epsilon$
the lattice spacing, in the following set to 1. Furthermore $a^\mu(x)= s^*_{\alpha}(x)\partial^\mu s_\alpha(x)$, with $\mu$ a space-time index, and the implicit constraint $s^*_{\alpha} s_\alpha=1$ is understood. However, in 1D there is an additional $\Theta=\pi$ term
\begin{eqnarray}
\label{eq11}
    \mathcal{L}_\Theta= i \frac{\Theta}{2 \pi} \epsilon_ {\mu\nu}\partial^\mu a^\nu.
\end{eqnarray}
(A hedgehog gas discussed in  Ref.\cite{sac} would appear in 2D instead of the $\Theta$-term, but it disappears when we add holons and therefore here will not be considered.)
Here we see at a deeper level the difference between 2D and 1D discussed at the beginning of this section: precisely at the value $\Theta= \pi$ in 1D there are two degenerate ground states connected by a parity transformation (see e.g. Ref. \cite{aff}) as in the Heisenberg chain. Kinks interpolating between the two destroy the Coulomb attraction generated by $a^\mu$, which is massless since the SU(2) symmetry is unbroken, and the spinons are deconfined. The absence of this $\Theta$-term in 2D makes the slave-particle gauge field confining in the phase with unbroken symmetry \cite{ien} and as a result the spinon $s$ and the anti-spinon $s^*$ are bound together into a spin 1 spin-wave. Also in the broken-symmetry antiferromagnetic phase in 2D the excitations are spin waves, but now they appear as Goldstone bosons. When we introduce holons, as we will see, we go from long-range to short-range AF in the $CP^1$ model, and in this approach the main actors are the antiferromagnetic spin-vortices,  at last! Confinement is destroyed by the vacuum polarization of holons, but there remains a short-range attraction between holon and spinon and between spinon and anti-spinon mediated by $a$ which has no analogue in 1D. In fact, due to the finite density of holons, the propagator of the transverse component of the slave-particle gauge field $a_\perp$, absent in 1D, is of the form

\begin{eqnarray}
\label{eq12}
   \langle a_\perp a_\perp \rangle (\omega,\vec k) \sim (- \chi |\vec k |^2 + i
   \kappa {\omega\over |\vec k|})^{-1},
\end{eqnarray}
for $\omega, |\vec k|, \omega/|\vec k| \sim 0$, 
where $\chi$ is the diamagnetic susceptibility and $\kappa$ the Landau damping.
By plugging as typical energy scale $T$ we see that there is a typical momentum scale, the so-called Reizer \cite{rei} momentum, $Q(T) = (\kappa T/\chi)^{1/3}$.
Below this momentum scale, or alternatively beyond the inverse momentum scale in space, there is an effective attraction mediated by $a_\perp$ between opposite charges relative to the slave-particle $U(1)$ group. Since we are in 2D we expect that an attractive interaction binds together excitations with opposite charges. In fact a weak binding is precisely the result obtained in 2D within some approximations in Refs. \cite{msy00}, \cite{msy04} for the hole as composite of spinon and holon and for the magnon as a composite of spinon and anti-spinon. They exhibit a ``small'' resonance life-time, $T$-dependent, originated from the slave-particle gauge coupling, leading to a behaviour of
these excitations less coherent than in a standard Fermi-liquid. For some considerations supporting a  gauge-induced composite structure
of the hole, see Ref. \cite{ay}.

\section{The charge- and antiferromagnetic spin-vortices}\label{aba:sec6}

If we accept the suggestion coming from the 1D model for the statistics of holon and spinon, one should search for a semionic representation of the hole field also in 2D. However in 2D the typical topological excitation is the vortex, not the kink, characteristic of 1D. Therefore the analogy with 1D suggests the appearance of a charge-vortex attached to the holon and a spin-vortex attached to the spinon and this is indeed what happens.
These vortices are somewhat analogous to those introduced by Laughlin in
the FQHE and in fact a semionic representation of the hole was advocated by him \cite{La} quite soon after the discovery of high $T_c$.

We now make these ideas precise in  the spin-charge gauge formalism.
Let us come back to the action of the 2D gauged $t$-$J$ model in the form of (\ref{eq4}). As in 1D one can integrate out $B^0$; we obtain the constraint (with $\mu,\nu=1,2, z \in \mathbf{R}^3$):
\begin{eqnarray}
\label{eq13}
\epsilon_{\mu\nu}\partial^\mu B^\nu(z)=\pi [\sum_j (1-h^*_jh_j)(z^0)\delta^{(2)}(\vec z-j)].
\end{eqnarray}
Imposing the Coulomb gauge-fixing $\partial_\mu B^\mu=0$ one gets $B^\mu(z)=\bar B^\mu + b^\mu(z)$ where $\bar B^\mu$ introduces a $\pi$-flux phase, i.e. $\exp[\int_{\partial p} \bar B]=-1$ for every plaquette $p$ and
\begin{eqnarray}
\label{eq14}
b^\mu(z)=-\frac{1}{2}[\sum_j \partial^\mu \arg(\vec z-j)( h^*_jh_j)(z^0))].
\end{eqnarray}
 We can recognize $ \partial^\mu \arg(\vec z-j)$ as the vector  potential of a vortex centered on the holon position $j$, i.e. centered on an empty site of the $t$-$J$ model. We already see here that strings in 1D are replaced in 2D by vortices.  The vortices in (\ref{eq14}) appear in the charge $U(1)$ group but their spin companions will be finally the antiferromagnetic vortices we are looking for the pairing. Following an argument given in Ref. \cite{ym} we conjecture that also in 2D, due to the $\pi$-flux,  the holon field $ \exp[i \int_{\gamma_j} b]h_j$ obeys  an exclusion statistics with parameter 1/2. A complete proof of this conjecture is still lacking but is presently under investigation and preliminary results are encouraging. In the following we take this conjecture for granted.
 
 We turn to the spin degrees of freedom and proceed as in 1D: we rewrite  $V^\mu= g^\dagger \bar V^\mu g+g^\dagger\partial^\mu g, \mu=0,1,2, g \in SU(2)$ with $\bar V$ satisfying the Coulomb condition $\partial_\nu \bar V^\nu=0, \nu=1,2$.
Then ideally one would search a configuration of $g$, depending on the holon configuration, optimizing the holon-partition function in that $g$ background, and around that configuration, $g^m$, one would consider spinon fluctuations. We didn't succeeded to find rigorously such configuration as in 1D, but we still found a configuration optimal ``on average'' by expanding the holon partition function in terms of holon worldlines in the first-quantization formalism as done in in one-dimension.
 Looking at (\ref{eq4}) we see that on links appearing in the $J$-term, thus not containing holons, the optimization requires $|(\sigma_x^{|i|} P(\exp[i \int_{\langle i,j \rangle}  V])\sigma_x^{|j|})_{11}|=0$. In the links of the hopping term the holons are coupled to the complex abelian gauge field:
\begin{eqnarray}
\label{eq15}
X_{\langle i,j \rangle}  =  \exp[i \int_{\langle i,j \rangle} B] [(\sigma_x^{|i|} P(\exp[i \int_{\langle i,j \rangle}  V])\sigma_x^{|j|})_{11}].
\end{eqnarray}
In Ref. \cite{msy98} it is shown that the optimal modulus of $X$ is 1 and
Lieb \cite{lie} has proved rigorously that at half-filling $(\delta = 0)$ the optimal
configuration for an abelian gauge field on a square lattice in 2D has a flux $\pi$ per plaquette at arbitrary temperature. On the basis also of the results of Refs. \cite{pi} and of a numerical simulation (L. Qin, unpublished) we conjecture that at least on average the optimal flux per plaquette is $\pi(1-\delta)$ for sufficiently small doping and temperature. On the other hand, it is well known that at high doping or temperature a vanishing flux per plaquette is favored. Therefore we expect a crossover corresponding to the melting of the $\pi$-flux lattice and we propose to identify such crossover, which we denote by $T^*$, with the experimental ``low-pseudogap'' crossover in the cuprates on the basis of comparison between theoretical curves and experimental data, in particular of in-plane resistivity as discussed in Sec. 10. Consistently with the terminology used in Sec. 2 for the phenomenology of the cuprates, we call PG also for the $t$-$J$ model in our approach the region below the $\pi \rightarrow 0$ crossover and we call SM the region above. We see from (\ref{eq6}) that the term $\exp[i \int_{\langle i,j \rangle} B]$ has in average a flux $\pi(1-\delta)$ per plaquette, hence in PG on the links derived from the hopping term we impose $(\sigma_x^{|i|}g_j^\dagger P(\exp[i \int_{\langle i,j \rangle}  \bar V]) g_i \sigma_x^{|j|})_{11}=1$. We find that $g^m$ involves a spin flip at the holon position as in 1D.  Hence writing $g= U g^m$ with $U$ as in (\ref{eq9}), the way in which the $s$ spinons appear in the $t$ and $J$ terms is the same seen in 1D . Therefore the considerations on the simultaneous optimization both of the $t$- and the $J$-term made at the end of sect. 4 apply, in approximate form,  also to 2D and this motivates ``a priori'' this spin-charge gauge approach. Recalling that in the continuum limit the lattice field $s$ generates its continuum version as discussed in the previous section, one realizes  that $s^*_{\alpha i}s_{\alpha j}$ appearing in the $t$-term of (\ref{eq4}) generates in the continuum limit a coupling of the holon with the slave-particle gauge field $a^\mu \approx s^*_\alpha \partial ^\mu s_\alpha$. In SM the optimal configuration $g^m$ contains also a phase factor deleting the contribution of $\bar B$ in the loops of hopping links of the holons, in order to get effectively an approximately 0 flux . Forgetting at first the spinon fluctuations and assuming an exclusion Haldane statistics with parameter 1/2 for the holon, in SM one then recover a ``large FS'' roughly consistent with band calculations \cite{msy05}. The effect of the field $\bar B$ is then to introduce an unbalanced $\pi$-flux in PG. As discussed in the slave-boson approach in Ref. \cite{lee}, through Hofstadter mechanism the $\pi$-flux converts the holon  of SM with dispersion, in this approximation,
$\omega_h \sim 2t[(\cos k_x + \cos k_y) - \delta]$ 
 into a pair of ``Dirac fields'', with pseudospin index corresponding to the two  N\'{e}el sublattices and
with dispersion:
$\omega_h \sim 2t [\sqrt{\cos^2 k_x + \cos^2 k_y }- \delta]$ 
restricted to the magnetic Brillouin zone (BZ). One thus obtain two ``small FS'' centred at $(\pm \frac{\pi}{2},\pm \frac{\pi}{2})$
 with
Fermi momenta $k_F \sim \delta$. When the holon is combined with the spinon to reconstruct the hole with dispersion in the full BZ, the Dirac structure of the holon suppresses the spectral weight outside of the magnetic BZ \cite{msy04}. 
Neglecting, as in 1D, the spinon fluctuations $U$ in the computation of $\bar V$ , one gets ( with $\mu=1,2)$:
\begin{eqnarray}
\label{eq16}
\bar V^\mu(z)= -\frac{1}{2} \sum_j (-1)^{|j|} \partial^\mu\arg(\vec z-j) h^*_jh_j(z^0) \sigma_z.
\end{eqnarray}
We recognize in the term $(-1)^{|j|} \partial^\mu\arg(\vec z-j)$ the vector potential of a vortex centered on the holon position $j$, with opposite vorticity (or chirality) for the center in opposite N\'{e}el sublattices. These are finally the antiferromagnetic spin vortices alluded in the beginning of the paper. As one see they are along the spin direction, $z$, of the magnetization; they are the topological excitations of the $U(1)$ subgroup of the $SU(2)$ spin group  unbroken in the antiferromagnetic phase. These vortices are of purely quantum origin, since of Aharonov-Bohm type, inducing a topological effect far away from the position of the holon itself, where their classically visible field strength is supported. Hence in this approach the empty sites of the 2D $t$-$J$ 
model, mimicking the Zhang-Rice singlets and corresponding to the holon locations, are the cores of spin vortices, a quantum distortion of the antiferromagnetic  spin background, recording in their vorticity the N\'{e}el structure of the lattice. Therefore they are still a peculiar manifestation of the antiferromagnetic interaction, like the more standard antiferromagnetic spin waves.
We now show that these vortices are responsible both for short-range AF and for charge-pairing.

\section{Short-range AF and charge pairing}\label{aba:sec7}

The effect of the spin vortices appeared in the last section is twofold: acting as impurities on the gapless spin waves of the Heisenberg model describing undoped cuprates they ``localize'' them, more precisely they make them gapped. Furthermore being a 2D gas of vortices with both chiralities ( corresponding to centers in the two N\'{e}el sublattices) they undergo a Berezinski-Kosterlitz-Thouless (BKT) transition. Since they are centered on holons the KT attractive pairing induces a charge pairing.

 Mathematically the presence of the antiferromagnetic spin vortices is recorded in the field $\bar V$ of (\ref{eq16}). In the low-energy continuum limit one can see from (\ref{eq4}) that in the $t$-term of the action $\bar V$ appears linearly; since its spatial average vanishes in a ``mean field'' treatment we ignore it. In the $J$-term instead it appears quadratically and assuming self-consistently that for sufficiently large $\delta$ the full $SU(2)$ symmetry is restored at large scales, treating the spinons in ``mean field'' one finds an interaction between vortices and spinons proportional to:
\begin{eqnarray}
\label{eq17}
\int d^3x (\bar V^\mu \bar V_\mu)(x) s^*_\alpha s_\alpha (x).
\end{eqnarray}
A quenched average, $\langle \cdot \rangle$, over the positions of the center of spin-vortices yields the following estimate  \cite{msy98}:  $\langle \bar V^\mu \bar V_\mu\rangle \approx \delta |\log \delta|$. Thus the term (\ref{eq17}) provides a mass-gap to the spinons. Leaving aside for the moment the monomial quartic in the holons of the $J$-term in (\ref{eq4}), we treat in mean field the monomial quadratic in the holons, so that the antiferromagnetic coupling is renormalized to $J (1-2 \delta)$. We see here the effect of strong reduction of antiferromagnetism due to the increase of the density of empty sites, corresponding to Zhang-Rice singlets in the cuprates. As a result one finds that the continuum limit of the term 

\begin{eqnarray}
\label{eq18}
\int dx^0 \sum_j i( \sigma_x^{|j|} V_0(x^0,j)  \sigma_x^{|j|})_{11}+ \nonumber\\ \sum_{\langle i,j \rangle} \frac{J}{2}(1-h^*_jh_j-h^*_ih_i) [|(\sigma_x^{|i|} P(\exp[i \int_{\langle i,j \rangle}  V])\sigma_x^{|j|})_{11}|^2-\frac{1}{2} ](x^0)
\end{eqnarray}
within the above approximations is given by
\begin{eqnarray}
\label{eq19}
\int d^3x (1-2 \delta)\frac{1}{g}
   \left[ v_{s}^{-2} \left| \left( \partial_{0} - i a_{0} \right) s_{\alpha}
   \right|^{2} + \left| \left( \vec \nabla - i \vec a
   \right) s_{\alpha} \right|^{2} 
   +m_s^2 s^*_{\alpha}s_{\alpha}
   \right](x),
   \end{eqnarray}
   where $m_s\approx\sqrt{\delta |\log \delta|}$.
 Hence the  gapless spinons $s$ forming the spin waves of the $CP^1$ (or equivalently $O(3)$) model describing the undoped system, traveling in a gas of spin vortices centered on holons acquire a gap $\approx ˜J (1-2 \delta)\sqrt{\delta |\log \delta|}$, converting the long-range AF of the undoped model in the short-range AF when doping exceeds a critical value. This is selfconsistent with the previous assumption of $SU(2)$ symmetry restoration at large scales. We thus see that in this approach the  transition  corresponding to the onset of short-range AF is indeed due to the antiferromagnetic spin vortices.
 We now show that the same term (\ref{eq17}) describing the interaction of spinons with spin vortices generates also the charge pairing, by treating in mean-field $ s^*_\alpha s_\alpha (x)$ instead of $\bar V^\mu \bar V_\mu(x)$.
 This averaging produces the term
 \begin{eqnarray}
 \label{eq20}
< s^*_\alpha s_\alpha > \sum_{i,j} (-1)^{|i|+|j|}\Delta^{-1} (i - j) h^*_ih_i h^*_jh_j , 
  \end{eqnarray}
  where $\Delta$ is the 2D Laplacian. In the static approximation for holons (\ref{eq20}) describes a 2D lattice Coulomb gas with charges $\pm 1$ depending on the
N\'{e}el sublattice. In particular the interaction is attractive between holons in opposite N\'{e}el sublattices, with maximal strength for nearest neighbour sites, along the lattice directions with a $d$-wave symmetry. Putting back the coefficients one finds that the coupling constant of this interaction is $J_{eff}=J(1-2\delta)< s^*_\alpha s_\alpha >$, thus decreasing with doping.  For
2D Coulomb gases with the above parameters, pairing appears
below a temperature $T_{ph} ˜\sim J_{eff}$ which turns out to be
inside the SM ``phase'' . Comparing the effect of this pairing on the spectral weight of the hole with ARPES data we propose to identify this crossover temperature with the experimental  ``high'' pseudogap in the cuprates, discussed in Sec.2. The formation of holon pairs, in fact, induces a reduction of the
spectral weight of the hole, starting from the antinodal region, which will be discussed in Sec.9.  Inserting the nnn hopping term, $T_{ph}$ strongly depends on its coefficient $t'$, contrary to the ``low'' pseudogap $T^*$. To identify more precisely the crossover temperature $T_{ph}$ we use a continuum approximation for the effective interaction, valid in the large-scale
limit, taking into account the Coulomb screening effect,
with a screening length $\ell \sim 1/\sqrt{k_F}$  in the Thomas-Fermi
approximation, where $k_F$ is the average fermi momenta of the FS of the holon. Then we treat the resulting Coulomb-screened attractive interaction in the BCS approximation obtaining a $d$-wave order parameter for the holons \cite{mfsy}\cite{mg}. The $d$-wave nature of the order parameter is a natural consequence of the $D_2$ (or $Z_2 \times Z_2$) symmetry of the FS of holons due to the ``charge'' $\pi$ flux and the structure of the holon pairing  (\ref{eq20}), both arising from AF . The modulus of the order parameter, $\Delta^h$, on the FS turn out to be $\sim  J_{eff} \sqrt k_F \exp [-{\rm const.} ~t /J_{eff}]$ . As expected the holon-pair order parameter decreases with doping and its scale is proportional to $J$ and not to $t$, as natural due to its magnetic origin, but reduced by the effect of the empty sites. In the next section we show that this charge pairing, due to the antiferromagnetic spin vortices, indeed leads to superconductivity.

\section{Spin pairing and superconductivity}\label{aba:sec8}

The charge pairing alone does not yet lead to hole-pairing,
since the spins are still unpaired. It is the slave-particle gauge attraction between holon and spinon that induces the formation of short-range
spin-singlet (RVB) spinon pairs at a lower temperature $T_{ps}$, in a sense, using the holon-pairs
as sources of attraction . 
More in detail:
the spinon gap is due to the presence of a gas of unpaired spin vortices and when the charge-pairing occurs the tight vortex-antivortex pairs don't contribute anymore to the spinon gap. Hence charge-pairing leads to a lowering of the spinon energy proportional to the density of spinon pairs. Furthermore, the monomial quartic in the holons of the $J$ term of the action in (\ref{eq4}) which involves the RVB spinon-pair field is repulsive and its condensation energy is negative and  decreased by the formation of spinon pairs. However fortunately the lowering of the spinon gap implies an enhancement of the contribution of the vacuum energy of the slave-particle gauge field. The competition between these two effects finally produces a saddle point with finite density of RVB spinon-pairs. From the combined charge- and spin-pairs we get a gas of incoherent spin-singlet hole-pairs.
Finally, at a even lower temperature, the superconducting
transition temperature $T_c$, the hole pairs become coherent and a $d$-wave hole condensate (in BCS approximation) appears, leading to superconductivity.
The appearance of two temperatures, one for pair formation
and a lower one for pair condensation, is typical of a BEC-BCS
crossover regime for a fermion system with attractive interaction
\cite{no}. In this sense the incoherent hole pairs discussed above play a role analogous to that of the ``preformed pairs''
considered e.g. in Refs. \cite{pp}. However, due to the composite spin-charge structure of the hole we have here two distinct temperatures: for charge-pair formation, $T_{ph}$, and for spin-pair formation, $T_{ps}$.
Let us look how the above ideas are implemented mathematically.

As remarked above in this approach a key ingredient of the spinon pairing is the monomial quartic in the holons of the $J$ term in (\ref{eq4}) that we have so far ignored. Rewritten in terms of the spinon fluctuations $s$ and neglecting the subleading contribution of $\bar V$ this Heisenberg term  is given by: $\frac{J}{2}h^*_jh_jh^*_ih_i |\epsilon_{\alpha\beta} s_{\alpha i}s_{\beta j}|^2 $. This interaction term is clearly  positive, hence repulsive, due to the semionic mean-field approach, contrary to the similar term in the slave-boson approach. It can be reasonably neglected in the normal state, since in mean field it is of $O(\delta^2)$, but as soon as holon-pairing becomes relevant the quartic holon monomial may become in mean-field $O(\delta)$ and it cannot be neglected anymore. To investigate its effect we apply a Hubbard-Stratonovic (HS) transformation and go to the continuum limit. Denote in this limit the HS field by $\Delta_s^\mu, \mu=1,2$ and treat the holon pairs in mean field. Then the additional term that one obtains to add to the spinon lagrangian in (\ref{eq19}) takes the form $-2 |\Delta_s(x)^\mu|^2/(J |\langle h_jh_i\rangle|^2) + \Delta_s(x)^\mu \epsilon_{\alpha\beta} s_{\alpha}\partial_\mu s_{\beta}(x)+ h.c. $ Treating the HS field in mean-field one finds \cite{mfsy} that a non-vanishing expectation value, $\Delta_s$, of $\Delta_s(x)$ modifies the positive branch of the spinon dispersion, $\omega_s(\vec k) =  \sqrt{m_s^2 +
\vec k^2}$, replacing it with two branches:
\begin{eqnarray}
\label{eq21}
 \omega_{s\pm} (\vec k) =  \sqrt{(m_s^2 - |\Delta_s|^2) +
(|\vec k| \pm |\Delta_s|)^2.}
\end{eqnarray} 
This doubling arises because for a
finite density of spinon pairs there are two (positive energy)
excitations, with different energies, but the same spin and
momenta. They are given, e.g., by creating a spinon up in the presence of the pairs and
by destructing a spinon down in one of the RVB spin-pairs.  
The lower branch exhibits a minimum with an energy lower
than $m_s$, analogous to the one appearing in a plasma of relativistic
fermions \cite{we}; it implies a backflow of the gas of spinon-pairs
dressing the ``bare'' spinon. Hence RVB pairing 
lowers the spinon kinetic energy.
 However, to get a non-vanishing  $\Delta_s$ in mean-field we
needs to compete with the spinon repulsion
generated by the  Heisenberg term discussed above; after the HS transformation the corresponding relevant lagrangian term is $\sim-|\Delta_s|^2/ |\langle h_jh_i\rangle|^2$. For this competition the slave-particle gauge field  is the relevant actor. In fact, since the spinon is gapped in absence of RVB pairing its contribution to the gauge effective action is a Maxwell term of the form $(\partial^\mu a^\nu-\partial^\nu a^\mu)^2/m_s$. When spinon pairing occurs we see from (\ref{eq21}) that the mass is reduced to $ M_s \approx m_s-|\Delta_s|^2/m_s$
(for small $|\Delta_s|/m_s$), hence integrating over the gauge field we get a lagrangian  contribution $\sim|\Delta_s|^2/m_s^2$ from the ground state variation. If it overcomes the previous term  one finds a non-vanishing saddle point in $\Delta_s$. In the underdoped region this occurs only beyond a critical concentration, since $|\langle h_jh_i\rangle|^2 \sim O(\delta)$ but $m_s^2 \sim O(\delta |\log \delta|)$. More precisely, the resulting ``gap equation'' has the form:
\begin{eqnarray}
\label{eq22}
\frac{3}{2 m_s}-\frac{1}{||\langle h_ih_j\rangle|^2}=\frac{1}{2 |\Delta_s| V} \sum_{\vec k}[\frac{k}{\omega_{s-} \tanh\frac{\omega_{s -}}{2T}}-\frac{k}{\omega_{s +} \tanh\frac{\omega_{s +}}{2T}}].
\end{eqnarray}
As soon as the RVB pairs in mean-field
condense at  $T_{ps}$ the slave-particle global gauge symmetry is broken
from $U(1)$ to $Z_2$ in mean-field. The Anderson-Higgs mechanism then implies
a gap for the gauge field $a^\mu$. However if we reinsert the phase fluctuations of the holon-pair and spinon-pair order parameters, the true superconducting transition is shifted below  $T_{ps}$, which remains just as a crossover. In fact
below $T_{ps}$ the low-energy effective action obtained integrating
out spinons is a Maxwell-gauged 3D XY model,
where the angle-field $\phi$ of the XY model is the ``square root'' of the phase of the long-wave limit of the hole-pair
field and the gauge field is the slave-particle gauge field (suitably dressed).
The holon contribution is  subleading and will be ignored here. The euclidean effective action is then given by
\begin{eqnarray}
\label{eq23}
S(\phi,a_\mu) \approx  \frac{1}{6 \pi M_s}\int d^3x[(\partial_\mu a_\nu -\partial_\nu a_\mu)^2 + 2 |\Delta_s|^2 (\partial_0 \phi -a_0)^2 \nonumber\\+ |\Delta_s|^2 (\vec \nabla \phi- \vec a)^2]
\end{eqnarray}
where, since in the range considered
$v_s/T \sim J/T << 1$, we extended the integration
to the full Euclidean space-time $\mathbf{R}^3$, but retaining 
the temperature dependence of the coefficients in the action. The
gauged XY model has two phases: Coulomb and Higgs.
If the coefficient, $\sim  |\Delta_s|^2/M_s$, of the Anderson-Higgs mass term
for $a$ is sufficiently small, the
phase field $\phi$ fluctuates so strongly that actually no mass gap is generated. This is the Coulomb phase, where there is 
a plasma of standard slave-particle ``magnetic'' (not spin!) vortices-antivortices. In the presence of
a temperature gradient a perpendicular external magnetic field induces an imbalance between vortices and antivortices,
giving rise to a Nernst signal, even if the hole pairs are not condensed yet. Therefore we conjecture that $T_{ps}$
corresponds in the phase diagram of cuprates to the onset of non-superconducting diamagnetic and vortex Nernst
signals. Since it is dominated by spinons physics, this crossover is essentially independent of details of the Fermi surface, hence material-independent.
For a
sufficiently large coefficient of the mass term for $a$ the gauged XY model is
in the broken-symmetry phase: the fluctuations of $\phi$ are
exponentially suppressed and we have a quasi-condensation (power-law decaying order parameter) at $T > 0$; accordingly the ``magnetic'' vortex-antivortex
pairs become dilute, so the slave-particle gauge
field is gapped and hence suppressed in the long wavelength limit. At the same time the holon, and hence the hole, acquires the gap outside the nodes of the $d$-wave order parameter; this is the superconducting
phase.  Therefore, the superconducting
transition is almost of classical 3D XY type,
driven by condensation of  ``magnetic'' vortices and related
to phase coherence as in BEC systems, in spite of the
BCS-like charge-pairing discussed above.

\section{Phase fluctuations of charge-pairs and Fermi arcs} \label{aba:sec9}

When we reinsert the phase fluctuations of the charge-pair order parameter we are going beyond the BCS approximation. The main effect is to reproduce between 
$T_{ph}$ and $T_{c}$ the phenomenology of Fermi arcs coexisting with gap in the antinodal region discussed in Sec.2. 
Indeed when a gas of incoherent holon pairs is present, the scale of the inverse correlation length (``mass''  $m_{ph}$) of the quanta of the phase of the holon-pair field separates self-consistently low energy modes with a Fermi liquid  behavior from high energy modes with a $d$-wave superconducting behavior.  More precisely an approximate evaluation of  self-energy correction for the holon, $\Sigma$, due to these phase fluctuations in the longwavelength continuum limit gives \cite{mg}: 
\begin{eqnarray}
\label{eq24}
\Sigma(\omega, \vec k)= \frac{|\Delta_h(\vec k)|^2}{(i \omega +\omega_h(\vec k))}[1- \frac{m_{ph}}{\sqrt{\omega^2+\omega_h(\vec k)^2+m_{ph}^2}}],
\end{eqnarray}
where $\Delta_h(\vec k)$ is the $d$-wave holon-pair order parameter and $\omega_h(\vec k)$ the holon dispersion. Clearly if $m_{ph}=0$ one goes back to the BCS approximation and in the opposite limit $m_{ph} \rightarrow \infty$ the correction vanishes.
  The value of $m_{ph}$ decreases with $T$, thus 
 lowering $T$ one finds a gradual reduction of the
spectral weight on the FS at small frequency as we move away from the diagonals of the Brillouin zone, due to the $d$-wave structure of $\Delta_h(\vec k)$. Simultaneously at larger
frequencies we have the formation and increase of two peaks of intensity precursors of the excitations in the superconducting 
phase. The sketched mechanism of spectral weight
suppression on the FS exhibits the fingerprint of the presence of the slave-particle gauge field, because the smooth interpolation
between Fermi liquid and superconducting behaviour discussed above is actually due to the interaction of the phase of the holon pairs with the gauge
field \cite{mg}. Without the gauge field the superconducting-like peaks in the spectral weight in the normal phase are almost absent, as in BCS.
The physical hole is obtained as a holon-spinon bound state produced by the gauge attraction and it inherits the above holon
features, but with a strongly enhanced scattering rate, due to the spinon contribution. Notice that in PG the suppression of the spectral weight discussed above sums up with the antinodal gap produced by the ``charge''  $\pi$ flux, leaving only small isolated segments of FS for the hole. This seems in agreement with recent experiments \cite{se}.
Below $T_{ps}$ beyond mean-field we have to add the phase fluctuations also of the spin-pairs which combine with phase fluctuations of the charge-pairs to give rise to the hole-phase field $\phi$ introduced in the previous section. 

\section{Brief comparison with experiments } \label{aba:sec10}

In this final section we compare some experimental data with theoretical computations performed within the above sketched approach to the $t$-($t'$)-$J$ model, with some experimental inputs for one doping to fix parameters that are then used consistently in all the calculations. So one may verify that many experimental features of the low-energy physics of hole-doped cuprates can find a natural explanation within the formalism presented here.
\begin{itemize}
\item Phase diagram. In Fig.1 we present the $\delta$-$T$ phase diagram theoretically derived \cite{uni} compared with some experimental data. We find that the general structure of the experimental crossovers is well reproduced, with even a semi quantitative agreement for $T^*$ with the "low pseudogap" and $T_{ps}$ with the boundary of ``Nernst phase'' . The only crossover that is completely missed is the onset of charge-density waves around $\delta \approx 1/8$; we conjecture that this is due to a contribution of the oxygen orbitals not taken into account by the $t$-($t'$)-$J$ model.
\def\figsubcap#1{\par\noindent\centering\footnotesize(#1)}
\begin{figure}[!htbp]
\begin{center}
  \parbox{2.2in}{\includegraphics[width=2.1in]{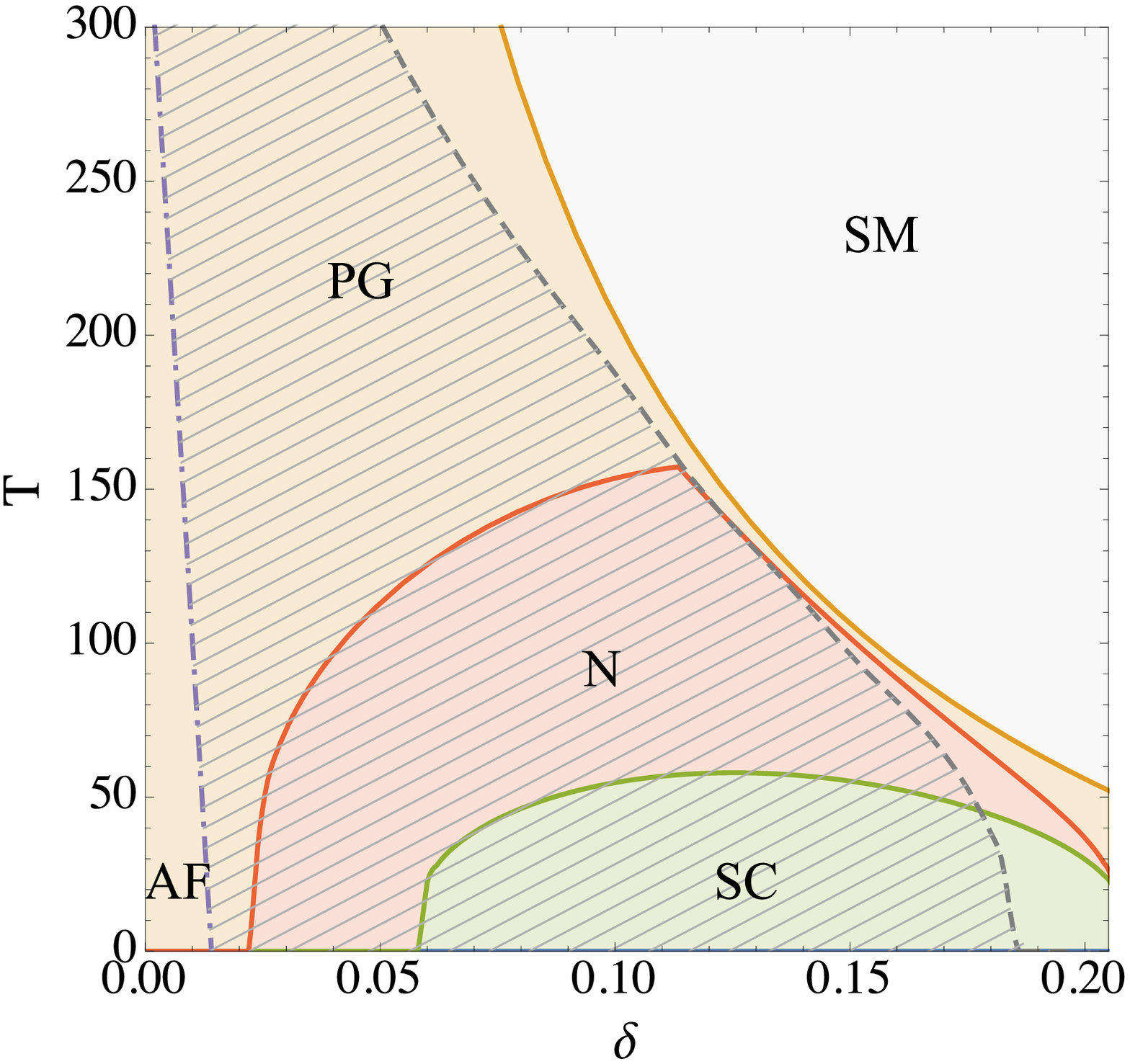}\figsubcap{a}}
  \hspace*{4pt}
  \parbox{2.1in}{\includegraphics[width=2in]{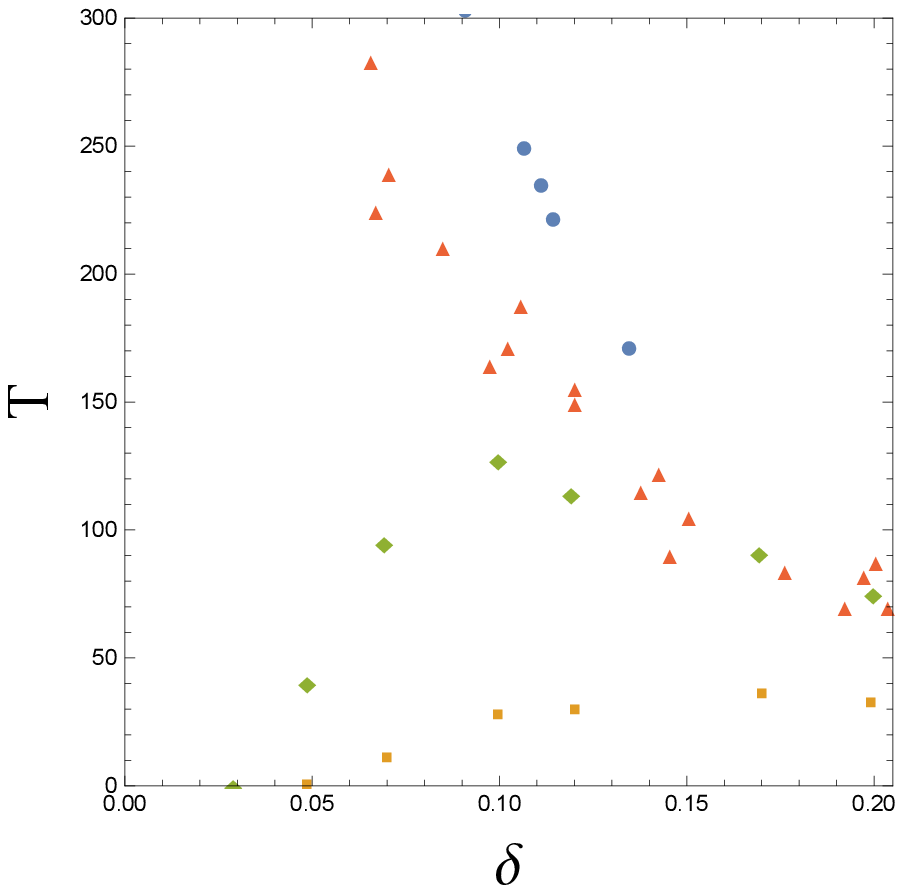}\figsubcap{b}}
  \caption{\footnotesize (a) Theoretically derived phase diagram from Ref. \cite{uni}: holon pairing temperature $T_{ph}$ (yellow line) determined from the BCS gap equation for holons, spinon pairing temperature $T_{ps}$ (red line) determined from the spinon gap equation, crossover PG-SM (dashed line) determined above $T_{ps}$ from the inflection point of resistivity and below $T_{ps}$ from matching the contour lines of the values of the spinon order parameter, $T_c$ (green line) determined from the transition temperature of the XY model of spinons. The N\`eel temperature (dot-dashed line) is qualitative from experiments not derived theoretically.(b) Experimental data for $T_c$ (yellow squares) and onset of Nernst signal (green diamonds) in LSCO from Ref.\cite{liwan}, ``low pseudogap'' (red triangles) in LSCO from Ref.\cite{ho} and ``high pseudogap'' (blue circles) in YBCO from Ref.\cite{ba}.}%
\end{center}
\end{figure}

\item In-plane resistivity $\rho_\|$. Since the hole is composite, due to the slave-particle gauge ``string''  between holon and spinon the in-plane resistivity is dominated by the slower of the two excitations (Ioffe-Larkin rule \cite{il}).  The antiferromagnetic spin vortices generate a spinon gap, therefore the resistivity is dominated by the spinons. In an eikonal treatment the interaction of the gauge field $a$ in PG  originates a saddle point producing a scattering rate for the spinon, $\Gamma$, behaving like $\Im \sqrt{m_s^2 + i c T/ k_F}$ with $k_F$ the Fermi momentum of the small holon FS \cite{msy04}. This in turn produces an inflection point of the resistivity at $T^* \sim m_s^2/k_F \sim |\log \delta|$. At a lower temperature, $T_{MIC}$, one finds also a metal-insulator crossover, intrinsic in this approach and not due to disorder \cite{msyprl}. It occurs when the insulating behaviour due to the spinon gap dominates, the time scale being set by spinon diffusion increasing with $T$, over the dissipative behaviour due to the gauge fluctuations. Furthermore, due to spinon dominance, the normalized resistivity defined by
 \begin{equation}
\label{eq25}
\rho_{\|n} (T) =\frac{\rho_\| (T)-\rho_\| (T_{\text{MIC}})}{\rho_\| (T^*)-\rho_\| (T_{\text{MIC}})} 
\end{equation}
exhibits an (approximate) universal behaviour \cite{msy04}\cite{uni} when expressed as a function of the normalized temperature $T/T^*$, in agreement with experiments as discussed in Refs. \cite{wuyts}. A comparison between experiments  and theory is shown in Fig. 2a.  $T^*$ appears also as an inflection point in the spin-lattice relaxation rate of the Cu ${}^63(1/T_1 T)$ (see e.g. Ref.\cite{be}) in PG and is reproduced in the above approach, see Ref. \cite{msy00}. In SM there is no saddle point in $\Gamma$ and one recovers \cite{msy05} the experimental resistivity linear in $T$ and the almost constant ${}^63(1/T_1)$.

\item Nernst signal. 
One expects that the amplitude of the spin-pair order parameter $\Delta_s$
is roughly proportional to the intensity of the Nernst signal and a comparison of the equi-level curves for $\Delta_s$ obtained solving the gap equation (\ref{eq22}) reasonably well compare with the intensity of the Nernst signal \cite{liwan}, as shown in Fig.2b.

\def\figsubcap#1{\par\noindent\centering\footnotesize(#1)}
\begin{figure}[!htbp]
\begin{center}
  \parbox{2.1in}{\includegraphics[width=2in]{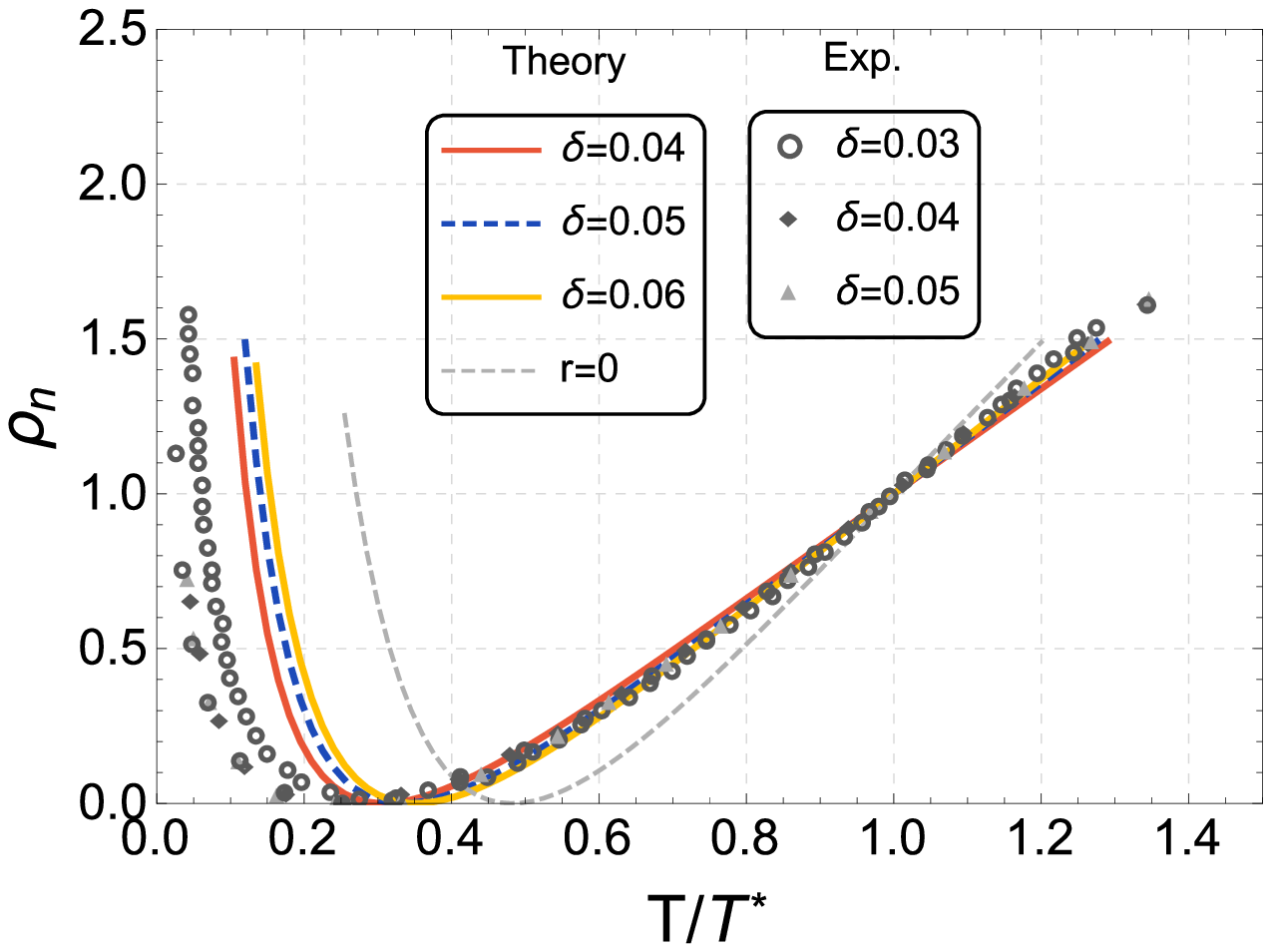}\figsubcap{a}}
  \hspace*{4pt}
  \parbox{2.1in}{\includegraphics[width=2in]{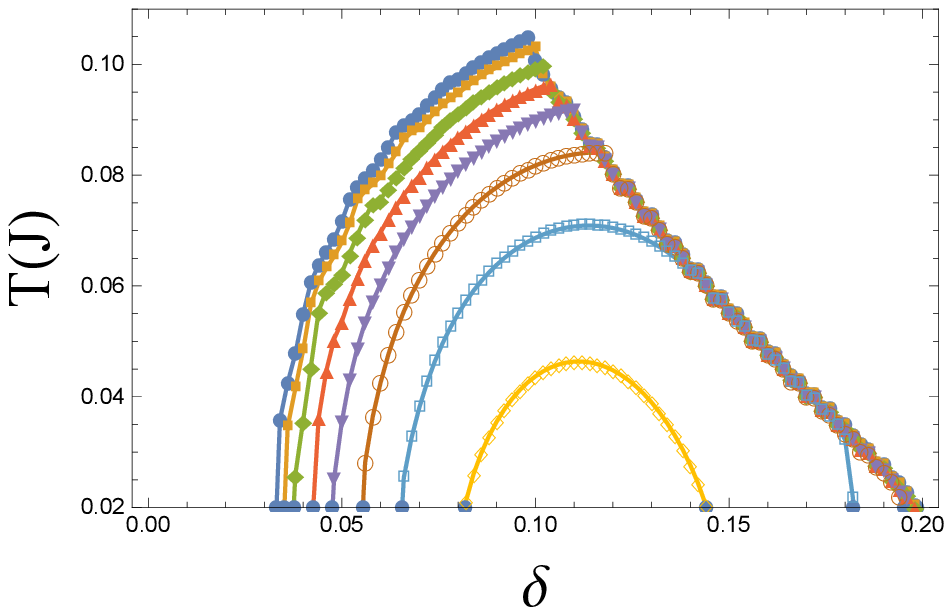}\figsubcap{b}}
  \caption{\footnotesize (a) The normalized in-plane resistivity $\rho_{\|n}$ theoretically calculated, including a holon contribution with relative weight $r$ w.r.t. the spinon contribution, from Ref. \cite{uni}, compared with experimental data from Ref. \cite{ando}. The $r=0$ curve corresponds to the universal, pure spinon contribution \cite{msy04}. The discrepancy at low $T$ might be due to the missing account of holon and spinon pair formation in the above treatment.(b) Equi-level lines of $\Delta_s$ from 0.19 to 0.26 in units of $J$ with $T$ in units of $J$; compare with experimental data in Ref.\cite{liwan}.}%
\end{center}
\end{figure}

\item Superfluid density $\rho^{(s)}$. 
For the same reason given for in-plane resistivity one finds that the superfluid density satisfies Ioffe-Larkin rule and in the underdoped region it is dominated by the spinons. From (\ref{eq23}), neglecting $a^\mu$ that is suppressed in the superconducting phase, one obtains a 3DXY model with effective temperature $\Theta(T) \sim M_s(T)/|\Delta_s|^2(T)$. Hence the superfluid density exhibits a 3D XY behaviour, with critical exponent 2/3, so definitely non-BCS, as already advocated for the experimental data in Ref.\cite{ca}. Furthermore $\rho^{(s)}$ normalized as
\begin{equation}
\label{eq26}
\rho^{(s)}_n (T/T_c) =\frac{\rho^{(s)} (T/T_c)}{\rho^{(s)} (T=0)} 
\end{equation}
 shows an universal behaviour \cite{mbi}, shown in Fig.3a, in agreement with this phenomenology noticed for a variety of materials in Ref. \cite{hardy}. Because $T_c$ is determined by the XY transition for the
effective temperature $\Theta(T)$ and $\Theta(0)=0$, expanding to the first order in $T$ one finds $T_c^{XY} =\Theta(T_c)\approx (d \Theta /dT)(0) T_c$, where $T_c^{XY}$ denotes the critical temperature of the XY model. But from the theory of the XY model we know that $\rho^{(s)}(0) \approx [(d \Theta /dT)(0)]^{-1}$, hence we derive an approximate Uemura relation \cite{ue}$\rho^{(s)}(0) \sim T_c$.

\item Spectral weight. The dependence on the
FS angle of the calculated symmetrized spectral weight of the hole is shown in Fig. 3b, it exhibits the Fermi arc phenomenology discussed in Sec.9  and it is compared with experimental data.

\def\figsubcap#1{\par\noindent\centering\footnotesize(#1)}
\begin{figure}[!htbp]
\begin{center}
  \parbox{2.1in}{\includegraphics[width=2in]{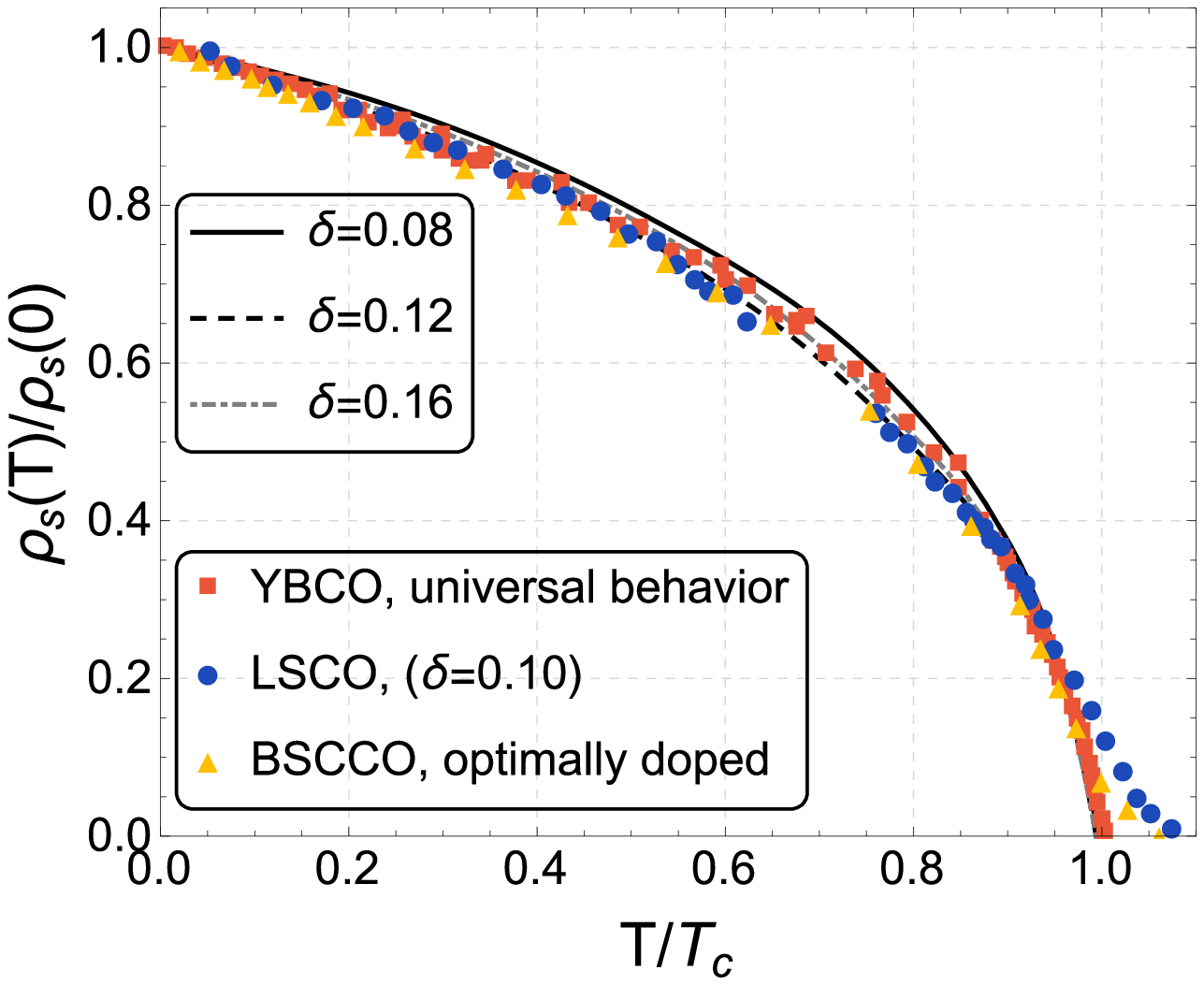}\figsubcap{a}}
  \hspace*{4pt}
  \parbox{2.1in}{\includegraphics[width=2in]{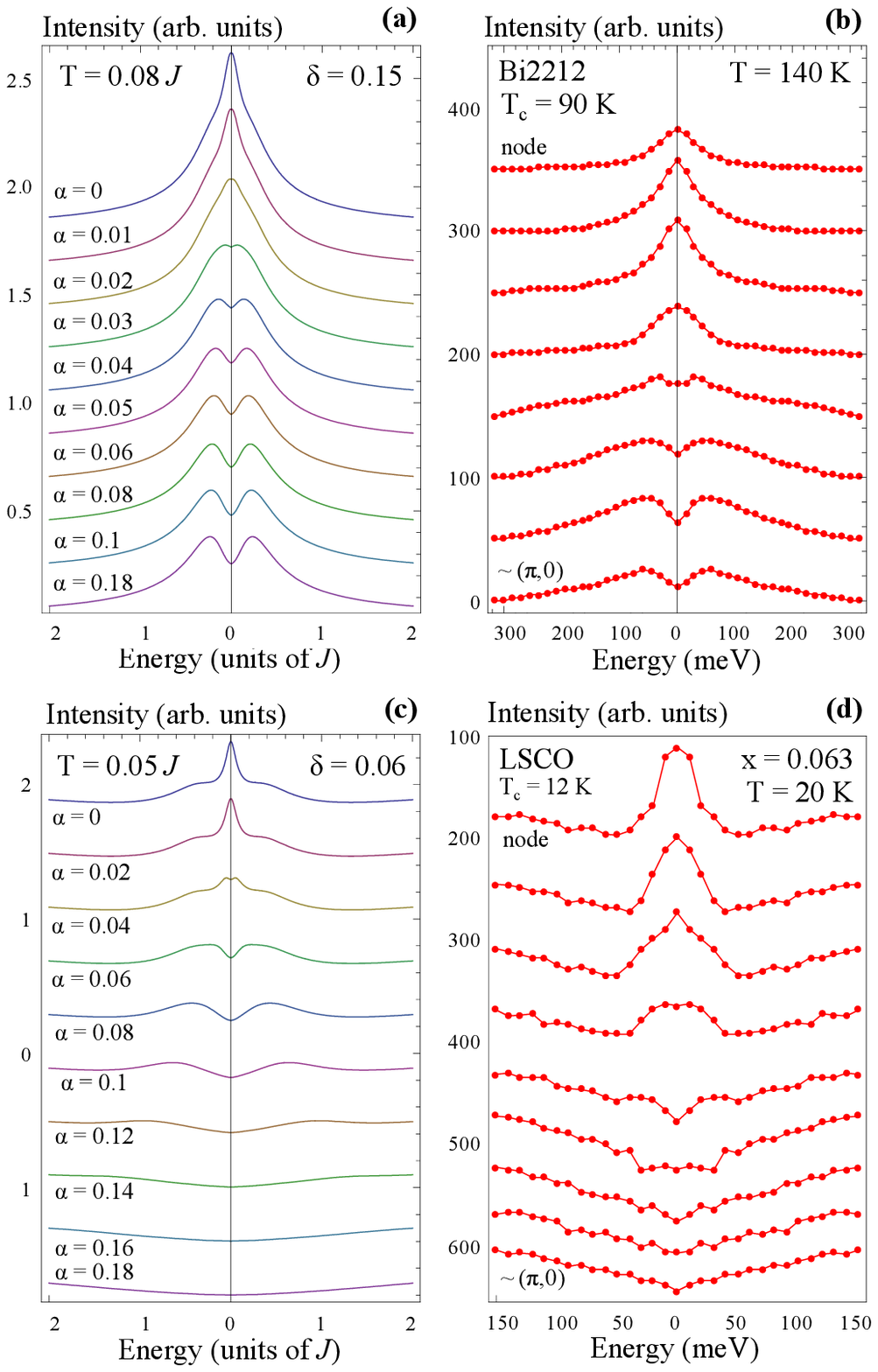}\figsubcap{b}}
  \caption{\footnotesize (a) The normalized superfluid density $\rho^{(s)}_n$ from Ref. \cite{uni}, compared with experimental data for underdoped (LSCO, YBCO) and optimally doped (BSCCO, YBCO) samples, from Refs. \cite{hardy}\cite{japa}. (b) Panel (a, c) shows the symmetrized spectral weight of the hole as a function of the FS angle $\alpha$ from the nodal direction  in SM and in PG, respectively, from Ref. \cite{mg}.
Panel (b) shows the experimental data for Bi2212 in Ref. \cite{ko} and panel (d) for LSCO in Ref. \cite{yo}.}%
\end{center}
\end{figure}

\end{itemize}

\vfill
\section*{References}



\end{document}